# Uniaxial Strain-Induced Electronic Properties Alteration of MoS$_2$ Monolayer

A Setiawan[1], I P Handayani[1, *], and E Suprayoga[2]

[1]Engineering Physics, School of Electrical Engineering, Telkom University, Bandung, Indonesia
[2]Research Center for Physics, National Research and Innovation Agency (BRIN), South Tangerang, Indonesia

*Email : iphandayani@telkomuniversity.ac.id

**Abstract.** Molybdenum disulfide (MoS$_2$) has attracted interest owing to its strain-tuned electronic and optical properties, making it a promising candidate for applications in strain engineering devices. In this study, we investigate the effect of uniaxial strain on the electronic properties of MoS$_2$ monolayer using first-principles calculations. Results show that a crossover of the K–K direct to –K indirect bandgap transitions occur at a strain of 1.743%. Moreover, a strong correlation is observed between the modified bandgap and the density of states (DOS) of the Mo-4$d$ and S-3$p$ orbitals at the valence band maximum and conduction band minimum. The uniaxial strain–tuned interatomic distance along the *a*-crystallographic axis does not only alter the bandgap at different rates but also affects the DOS of the Mo-4$d$ orbital and possible electronic transitions. This study clarifies the mechanism of the electronic structural modification of two-dimensional MoS$_2$ monolayer, which may affect intervalley transitions.

## 1. Introduction

Molybdenum disulfide (MoS$_2$) is a layered transition metal dichalcogenide with a honeycomb-like hexagonal structure comprising a layer of molybdenum sandwiched between two layers of sulfur [1, 2]. Exhibiting a direct bandgap of 1.8–1.9 eV at the **K** point [3–5], MoS$_2$ has attracted considerable interest owing to its nonzero bandgap and tunable electronic as well as optical properties, which can be achieved by varying the number of layers [6–8] and applying strain [3, 7, 9]. Experiments have shown that applied strain affects the electronic transport [10, 11], photoluminescence [12, 13], and direct-to-indirect bandgap transition [14] of MoS$_2$. Theoretical studies have revealed enhanced circularly polarized electroluminescence [15], the occurrence of valley drift, and decreased symmetry after the application of uniaxial strain [16]. These findings imply that elucidating the effect of strain on the electronic band structure of MoS$_2$ is important, particularly the alteration of electronic bands involving intervalley transitions. From the application viewpoint, strain engineering shows potential in tuning flexible-electronics properties [12, 17]. The reduced contact barrier increases the current and photoresponsivity of MoS$_2$ monolayer. In photonic devices, the strain-tuned bandgap, density of state (DOS) profile, and direct-to-indirect transition are promising for tunnel field-effect transistors (FETs) [18]. Direct bandgaps are suitable for FETs, whereas indirect bandgaps are employed in band-to-band tunneling processes. The photoresponsivity of photodetectors can also be tuned by varying the strain [19]. Moreover, the effects of strain on other systems, such as heterostructures [20], silicine [21], and silicane [22], have been discussed. Such strain effects are expected to benefit photocatalysts for water splitting, spintronic applications, and optoelectronic devices.

Density functional theory (DFT) calculations have been widely employed to elucidate the strain-induced intriguing properties of MoS$_2$, such as the valley drift [16, 23], electronic structures of wrinkled layers [24], and the effect of S vacancies [25]. These calculations have revealed the electronic structural modification of MoS$_2$ under various strains, which are crucial for understanding intervalley transitions and achieving potential strain engineering applications. Moreover, DFT calculations based on experimental data have been conducted to understand the observed experimental phenomena [15, 26]. Generally, direct-to-indirect bandgap transitions in MoS$_2$ occur when the applied strain ranges from 1% [4, 8, 10] to 2% [26] and are accompanied by the crossover of K–K direct to –K indirect [14, 26, 27]. Moreover, the effects of spin orbit coupling [18, 26] and DOS variations [18, 25] have been discussed to show the complex properties and promising applications of strain tuned MoS$_2$ properties.

To clarify the effects of strain on the electronic properties of MoS$_2$ monolayer in detail, we systemically investigate the change in the electronic band structure and DOS of MoS$_2$ monolayer under various applied strains along the *a*-crystallographic axis. The applied uniaxial strain not only decreases the direct and indirect bandgaps at different rates but also alters the DOS distribution between molybdenum and sulfur, thus playing an important role in interband electron transitions. In this study, we focus on the three *k*-points where the direct-to-indirect bandgap transitions occur. In particular, we investigate the projected DOS (PDOS) at the Γ point of the valence band maximum (VBM), K point of the VBM, and K point of the conduction band minimum (CBM).

## 2. Method

Herein, all calculations are performed based on DFT using the local density approximation (LDA) method with the Perdew–Zunger (PZ) functional for the exchange-correlation energy and implemented in Quantum ESPRESSO (QE) [28]. The LDA method does not consider the van der Waals interaction between layers [9] and has been widely implemented in MoS$_2$ monolayer systems [16, 23–25, 29].

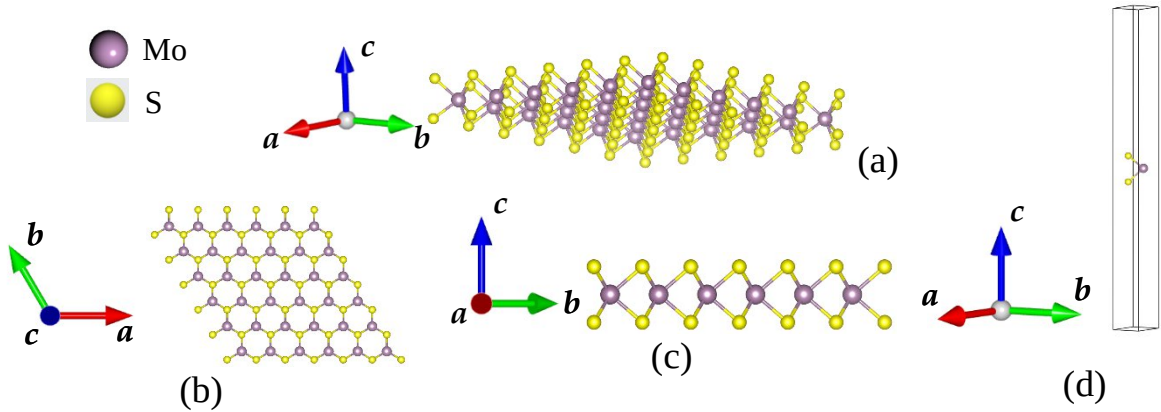

**Figure 1.** (a) MoS$_2$ monolayer structure. (b) Top view and (c) side view of a MoS$_2$ monolayer structure. (d) Unit cell of an optimized unstrained MoS$_2$ monolayer used for the calculation.

The initial structure of MoS$_2$ is optimized until the change in the total energy between two consecutive steps is less than $10^{-4}$ Ry and the components of all forces are smaller than $10^{-3}$ Ry/Å. The kinetic energy cutoffs for the wave function and kinetic energy of charge density are set to 70 and 700 Ry, respectively. The total energy of the system is kept constant for a greater kinetic energy cutoff, indicating that the selected kinetic energy cutoff is sufficiently high to converge the total energy. The Monkhorst–Pack method is used for Brillouin zone sampling with $k$- points of $30 \times 30 \times 1$ for lattice optimization and electronic structural calculations. Owing to the $xy$ plane orientation of the monolayers, the values of the Monkhorst–Pack set along the $x$ and $y$ directions are higher than that along the $z$ direction for the smooth sampling of the MoS$_2$ monolayer and time efficiency.

The initial position and lattice parameters are adopted from Materials Project [30]. A supercell is not used in this study because QE treats unit cells as crystals and for time efficiency. Figures 1(a)–(c) depict the crystal structure treated using QE. A vacuum space of ~18 Å is included along the *c* direction to eliminate the interlayer interactions. Thereafter, structural optimization is performed on the MoS$_2$ monolayer geometry to achieve the lowest energy. Figure 1(d) shows the unit cell of the optimized MoS$_2$ monolayer without the application of strain.

$$a = a_0 \, (1 + \varepsilon) \tag{1}$$

$$x = x_0 \, (1 + \varepsilon) \tag{2}$$

$$DOS_{\text{contr.}_{k,E,m_l}} = \frac{PDOS_{k,E,m_l}}{DOS_{total_{k,E}}} \tag{3}$$

Uniaxial strain is applied to the system using the MatSQ Structure Builder tool. The strain is applied along the *a*-crystallographic axis. The strain input $\varepsilon$ modifies the lattice parameter $a$ and atomic position $x$ because the *a*-crystallographic axis is parallel to the *x* axis. We use the method proposed by Dong *et al.* [5] and Nguyen *et al.* [31] to apply the strain. The modification of $a$ and $x$ using $\varepsilon$ is expressed in equations (1) and (2), where $a_0$ and $x_0$ are the initial lattice parameter and position, respectively. In this study, $\varepsilon$ is varied from 0%–2.8% with steps of 0.2%. The $y$ and $z$ components of the atomic positions are kept constant because the strain is applied along the *a*-crystallographic axis. Finally, the DOS contribution at specific $k$-points and energy levels ($k$, $E$) for specific orbital orientation ($m_l$) is calculated using equation (3).

## 3. Results and Discussion

### 3.1 The MoS$_2$ Crystal Structure

The crystal structure of the MoS$_2$ monolayer shows hexagonal graphene-like three-layer sheets. Mo atoms occupy the middle sublattice, and S atoms occupy other sublattices (Fig. 1(b)). The lattice parameter *a* of the optimized MoS$_2$ monolayer without the application of strain is determined to be 3.1660 Å, consistent with the experimental value of 3.20 Å [32].

Tables 1 and 2 list the lattice parameters of the optimized MoS$_2$ monolayer structure as well as those reported in other studies and show a comparison of the MoS$_2$ bond lengths. The lattice parameter *a* is smaller than the initial value of 3.190 Å and the experimental value of 3.20 Å [32]. This difference is attributed to the LDA method, which tends to overestimate the binding energy and yields smaller lattice constants [27]. Alternatively, the *c*-crystallographic axis shows a high value of 39.13 Å because of the application of the vacuum layer of ~18 Å to the system.

Using the generalized gradient approximation (GGA) method that utilizes the Perdew–Wang (PW91) functional, a lattice constant that is more consistent with the experimental results is achieved (Table 1). However, the GGA method generally underestimates the bandgap energy of semiconductors [10]. Consequently, the GGA method based on either the PW91 or Perdew–Burke–Ernzerhof (PBE) functionals, cannot appropriately describe the electronic properties of MoS$_2$. The bond length between molybdenum and sulfur and that between sulfurs are listed in Table 2. The calculation results agree well with the results obtained in earlier computational studies [8, 29].

### 3.2. MoS$_2$ Electronic Band Structure, DOS Contribution, and Strain-induced Direct-to-indirect Bandgap Transition

Figure 2 shows the electronic band structure, total DOS, and PDOS of Mo 4*d* orbitals and S 3*p* orbitals for each system under different applied strains. A direct bandgap is detected at the K point, and an indirect bandgap is observed from the Γ (VBM) to the K points (CBM). The Γ point of the VBM increases with a rate of 0.01175 eV/% with an increase in the applied strain. Alternatively, the K point of the VBM and CBM decreases at rates of 0.07138 eV/% and 0.1643 eV/%, respectively. The PDOS indicates the dominance of the Mo 4*d* orbital.

Figure 3 shows the indirect and direct bandgap shifts. The indirect bandgap decreasing rate (181 meV/%) is higher than the direct bandgap decreasing rate (95 meV); these values agree with those reported by Conley *et al.* [4]. The difference in the decreasing rates of the direct and indirect bandgaps induces a K–K direct to Γ–K indirect bandgap transition at a strain of 1.743% (the intersection of both lines in Fig. 3), which agrees with the experimental results [15, 26] and other computational findings [6, 27].

Figure 4 shows the DOS contribution evolution of the S-$3p$ and Mo-$4d$ orbitals as a function of increasing uniaxial strain. At the Γ point of the VBM, the DOS contributions are mainly from the Mo-$4d_{z^2}$ and S-$3p_z$ orbitals, whereas at the K point of the VBM, they are mainly from the Mo-$4d_{x^2-y^2}$, Mo-$4d_{xy}$, S-$3p_x$, and S-$3p_y$ orbitals. At the K point of the CBM, the DOS contributions are mainly from the Mo-$4d_{z^2}$, S-$3p_x$, and S-$3p_y$ orbitals. Because the DOS contributions from other orbital orientations are lower than 0.001%, they are not considered. The DOS contribution from the Mo-$4d$ orbital is higher than that from the S-$3p$ orbital for all the three points, indicating that the possibility of a $d$–$d$ transition is higher than that of a $p$–$d$ transition in the system. Because the direct transition occurs at the K point, we infer that the electrons tend to excite from the Mo-$4d_{x^2-y^2}$ and Mo-$4d_{xy}$ orbitals at the VBM to the Mo-$4d_{z^2}$ orbital at the CBM for MoS$_2$ monolayers without the application of strain. At strain higher than 1.743%, the indirect bandgap emerges, and the electrons are most likely excited from the Mo-$4d_{z^2}$ orbital of the VBM to the Mo-$4d_{z^2}$ orbital of the CBM.

Uniaxial strain increases the DOS contribution of the Mo-$4d$ orbital and decreases that of the S-$3p$ orbital. At the Γ point of the VBM, the DOS contribution of the Mo-$4d_{z^2}$ orbital increases at a rate of 0.728%$_{contr.}$/% with an increase in the applied uniaxial strain, whereas the DOS contribution from the S-$3p_z$ orbital decreases at a rate of 0.776%$_{contr.}$/%. At the K point of the VBM, the DOS contributions of the Mo-$4d_{x^2-y^2}$ and Mo-$4d_{xy}$ orbitals also increase at rates of 0.211%$_{contr.}$/% and 0.355%$_{contr.}$/%, respectively.

**Table 1.** Comparison of the calculated MoS$_2$ lattice parameters with experimental values (Exp.) and computational (Comp.) studies.

| | | | Lattice constant (Å) |
|---|---|---|---|
| Exp. | | | 3.20 [33] |
| | | | 3.27 [34] |
| Comp. | LDA | PZ | 3.17 (This study) |
| | GGA | PBE | 3.18 [25] |
| | | | 3.16 [35 – 38] |
| | | PW91 | 3.19 [8] |

**Table 2.** Comparison of MoS$_2$ bond lengths obtained herein with those obtained in other computational studies.

| Method | Functional | $d_{\text{Mo-S}}$ (Å) | $d_{\text{S-S}}$ (Å) |
|---|---|---|---|
| LDA | PZ (this study) | 2.42 | 3.17 |
| GGA | PW91 [10] | 2.43 | 3.17 |
| | PW91 [26] | 2.41 | 3.13 |

Moreover, the DOS contribution of the S-3$p_x$ orbital decreases at a rate of 0.434%$_{\text{contr.}}$/%, whereas that of the S-3$p_y$ orbital remains almost the same (decreases at a rate of 0.001%$_{\text{contr.}}$/%). At the K point of the CBM, the DOS contributions of the S-3$p_x$ and S-3$p_y$ orbitals decrease at rates of 0.292%$_{\text{contr.}}$/% and 0.486%$_{\text{contr.}}$/%, respectively, whereas that of the Mo-4$d_{z^2}$ orbital increases at a rate of 0.662%$_{\text{contr.}}$/%.

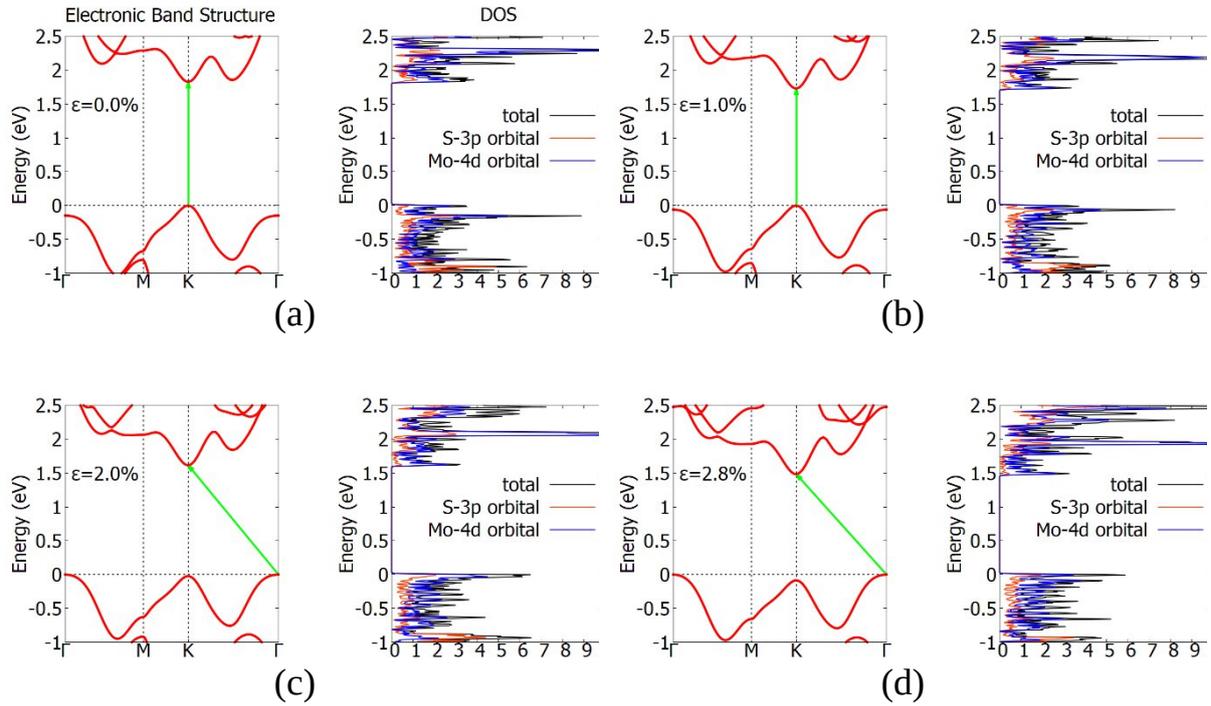

**Figure 2.** Electronic band structure, DOS, and PDOS of $MoS_2$ monolayer under a strain of (a) 0.0%, (b) 1.0%, (c) 2.0%, and (d) 2.8%. The energy axes are shifted to the Fermi level.

To understand the increase in the DOS contribution of the Mo-$4d$ orbital and decrease in that of the S-$3p$ orbital, we consider the Mo-S covalent bond and the electronegativity of both atoms. The molybdenum atom shares all six valence electrons with both sulfur atoms, and each sulfur atom shares two valence electrons with the Mo and other S atoms. Hence, the shared electrons are mainly from molybdenum. Furthermore, the molybdenum and sulfur atoms have the electronegativity of 2.16 and 2.58, respectively; thus, the electron density tends to be adjacent to the sulfur atoms. Based on this, the applied uniaxial strain along the *a*-crystallographic axis may shift the electron density toward the Mo atoms. Consequently, the DOS distribution of the Mo $4d$ orbital increases, whereas that of the S-$3p$ orbital decreases.

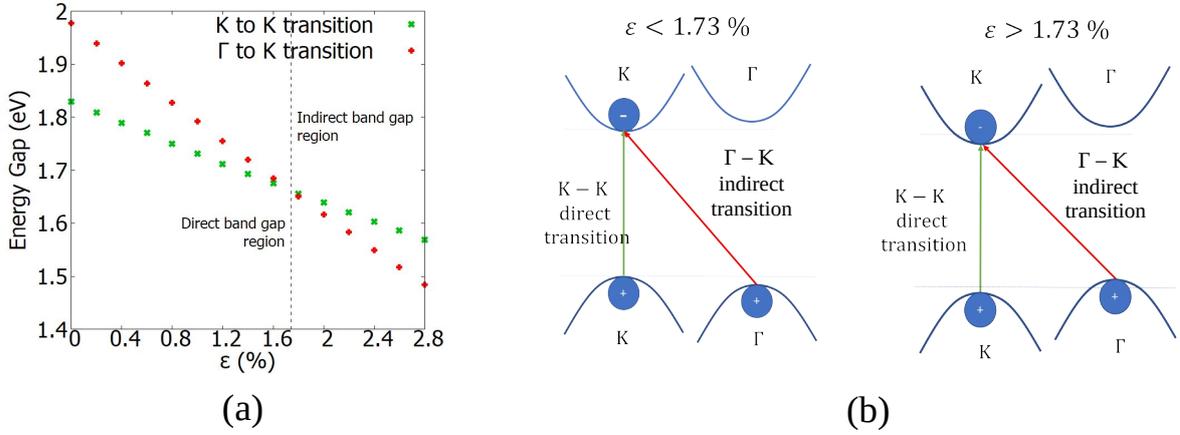

(a) Strain-induced direct and indirect bandgap shifts in MoS$_2$ monolayers. (b) Electronic band structure and transitions under various strains. The dashed lines show the VBM and CBM levels in the MoS$_2$ monolayer without the application of strain.

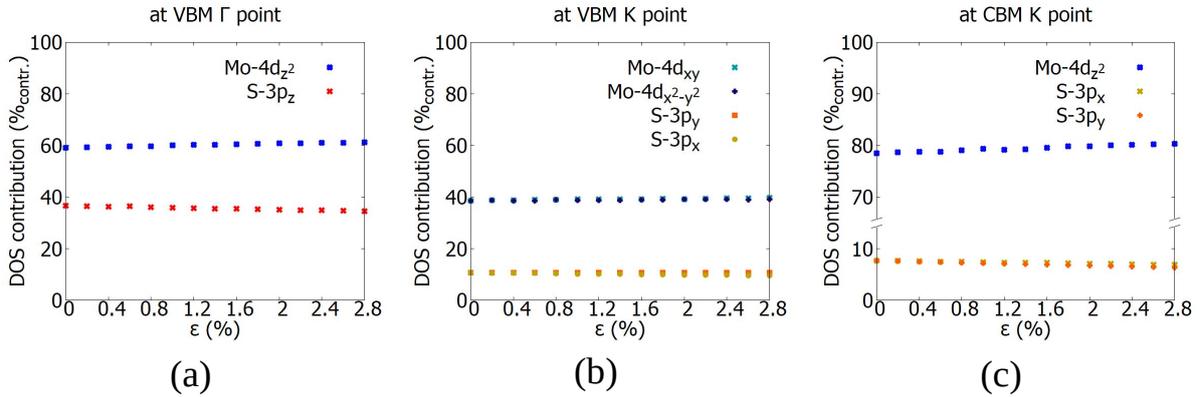

**Figure 4.** DOS contribution in a MoS$_2$ monolayer at the (a) $\Gamma$ point of the VBM, (b) K point of the VBM, and (c) K point of the CBM under applied strain.

We infer that an increase in the uniaxial strain along the *a* direction not only increases the lattice constants and bond lengths but also modifies the orbital occupations, which manifest on the DOS contribution. The increasing DOS contribution rate of the Mo- $4d_{z^2}$ orbital at the $\Gamma$ point of the VBM is higher ( $0.728\%_{\text{contr.}}/\%$) than the sum of the increasing DOS contribution rates of the Mo-$4d_{x^2-y^2}$ and Mo-$4d_{xy}$ orbitals at the K point of the VBM ($0.211\%_{\text{contr.}}/\% + 0.355\%_{\text{contr.}}/\% = 0.566\ \%_{\text{contr.}}/\%$). This phenomenon is understandable considering the *xy* plane orientation of the Mo-$4d_{x^2-y^2}$ and Mo $4d_{xy}$ orbitals;

moreover, the Mo-$4d_{z^2}$ orbital is at the center of the atom orientation. These different shifting rates between the DOS contributions at the Γ and K points of the VBM can change the possible electronic transition and result in a faster decrease in the indirect bandgap where the VBM level is located at the Γ point compared with the direct bandgap at the K point (Fig. 4). Compared with the direct bandgap, the faster decreasing rate of the indirect bandgap subsequently yields the direct-to-indirect bandgap transition.

## 4. CONCLUSION

First-principles calculations are performed on $MoS_2$ monolayer to investigate the effects of applied uniaxial strain on its electronic properties. Generally, the applied strains increase the Mo-*4d*- and decrease the S-*p*- DOS contributions. The crossover of the K − K direct to − K indirect transition is observed at a strain of 1.743%. This transition is strongly correlated with the faster-increasing rate of the DOS contribution of the Mo-$4d_{z^2}$ orbital at the Γ point of the VBM compared with those of the Mo-$4d_{x^2-y^2}$ and Mo-$4d_{xy}$ orbitals at the K point of the VBM. We investigated the mechanisms of the electronic band alteration in $MoS_2$ monolayer under applied strain. Thus, this study provides better insight into intervalley transitions, flexible electronics, and photonics devices. Further research is required to confirm the shift in the electron density and changes in the orbital orientation to clarify their role in intervalley transitions.

**Acknowledgments**

This work was financially supported by Direktorat Riset dan Pengabdian Masyarakat Direktorat Jendral Pendidikan Tinggi (DRPM DIKTI Kemenristek/BRIN) under Nos. 1867/E4/AK.04/2021, 065/E4.1/AK.04.PT/2021, 004/SP2H/RDPKR-MONO/LL4/2021, and 384/PNLT2/PPM/2021. The computation in this work has been done using the facilities of HPC BRIN, National Research and Innovation Agency (BRIN).